\documentclass[preprint,11pt]{elsarticle}

\usepackage[labelsep=period]{caption}

\usepackage[english]{babel}
\usepackage{graphicx}
\usepackage{geometry}
\usepackage[section]{placeins}
\usepackage{float,caption}
\usepackage{float}

 \usepackage{epsfig}

\usepackage{mathtools, bm}
\usepackage{amssymb}
\usepackage{graphicx}

  \usepackage{amsmath}
   \usepackage{amsfonts}
   \usepackage{amssymb}

\usepackage{comment}

\usepackage{color}
\usepackage{amsfonts}

\newcommand{\azd}{\textcolor{black}}

\usepackage{lineno}
\usepackage{amsmath}

\usepackage{soul}

\usepackage{algorithm}
\usepackage[noend]{algpseudocode}

\makeatletter
\def\BState{\State\hskip-\ALG@thistlm}
\makeatother

\makeatletter
\renewcommand\paragraph{\@startsection{paragraph}{1}{\z@}%
            {-2.5ex\@plus -1ex \@minus -.25ex}%
            {1.25ex \@plus .25ex}%
            {\normalfont\normalsize}}
            
\makeatother
\usepackage{ulem}

\begin{document}

\begin{frontmatter}

\selectlanguage{english}
\title{Non-local modeling with asymptotic expansion homogenization of randomly inclusions reinforced materials.}

\author{Sami Ben Elhaj Salah$^a$, Azdine Nait-Ali$^a$, Mikael Gueguen$^a$, Carole Nadot-Martin$^a$ \\}
\address{%
$^a$ Institut Pprime, UPR CNRS no 3346, CNRS – Universit\'e de Poitiers – ENSMA, Physics and Mechanics of Materials Department, ISAE-ENSMA, 1 avenue Cl\'ement Ader, BP 40109, 86961 Futuroscope, Chasseneuil, France \{sami.ben-elhaj-salah, azdine.nait-ali, mikael.gueguen, carole.nadot\}@ensma.fr\\
}

\address{\small Received *****; accepted after revision ++++\\
Presented by ++++}

\begin{abstract}
The aim of this study is to build a non-local homogenized model for three-dimensional composites with inclusions randomly embedded within a matrix according to a stochastic point process $w=(w_{i})_{i\in\mathbb{N}}$ in a bounded open set of $\mathbb{R}^{3}$ associated with a suitable probability space ($\Game, A, P$) as defined in \citep{nait2017nonlocal} and   \citep{michaille2011macroscopic}. Both phases were linear elastic. Asymptotic expansion homogenization (AEH) was revisited by taking into account the stochastic parameter $(w)$ representing the inclusion centers distribution. The macroscopic behavior was then studied by combining the variational approach with the mean-ergodicity. At the end, the advanced approach makes naturally emerge non-local terms (involving the second displacement gradient) as well as a strong microstructural content through the presence of the characteristic tensors in the expression of the homogenized elastic energy. Microstructures with a high contrast between constituents Young$'$s modulus leading to non-local effects were considered to test the model. Virtual microstructures were first generated with a fixed, simple, pattern before considering real microstructures of Ethylene Propylene Dien Monomer (EPDM) containing cavities in order to envision morphological situations with increasing complexity.

\end{abstract}


\end{frontmatter}

\clearpage
\newpage
\section{Introduction}
The best known non-local models have been developed in the past by \cite{kroner1967elasticity}, \cite{krumhansl1968some} and \cite{eringen1972nonlocal}. These models are based on the theory that the response of a material point depends on the deformation of this point as well as of its neighboring points. Non-local phenomena are much studied especially in the case of composites. Phenomenological models are constructed according to the composite morphology \cite{Pedro}. Non-local theory can be applied to materials with different behavior laws. In the case of isotropic linear elasticity, it is assumed that the strain gradient also plays a role in the material response in addition to the strain, see \cite{mindlin1968first}. In this reference, it is shown that the mathematical formulations based on the second gradient of the displacement or on the first gradient of the (symmetric) strain tensor are equivalent \citep{Aifantisa}. The dual quantities of the first and second gradients of the displacement field in the work density of internal forces, $W^i$, are the second rank simple force stress tensor ${\sigma}$ and the third rank double force stress tensor or hyperstress tensor ${S}$.

\begin{equation}
{W}^{i}(x)={\sigma}:\nabla {u}+{S} \vdots \nabla \nabla {u}
\end{equation}

In the case of damage, the localization process requires the regularization of the stress-softening term. A possible choice is to regularize the model through the introduction of a gradient term of the damage variable. In the variational approach of damage mechanics, such a regularization can be achieved in an elegant way by adding a non-local counterpart to the local part of the total energy which depends on the gradient of the damage variable. It is mandatory to introduce the parameter of internal length in such regularized damage models. In the model of Marigo et al. \cite{marigo2015gradient}, the energy density is the sum of three terms (see Eq. (\ref{eq marigo})): the stored elastic energy  $\psi(\epsilon,\alpha)$, the local part of the dissipated energy by damage $w(\alpha)$ and its non-local part $\frac{1}{2} w_{1}l^{2}.g.g$. The triplet ($\epsilon,\alpha,g$) denotes respectively the strain tensor, the damage parameter and the gradient vector of damage ($g=\nabla\alpha$).

\begin{equation}
{W}^{1}(\epsilon,\alpha,g)=\psi(\epsilon,\alpha)+w(\alpha)+\frac{1}{2}w_{1}l^{2}g.g
\label{eq marigo}
\end{equation}

Like all softening laws, the Mazars \cite{mazars1989strain} local model poses difficulties related to the phenomenon of deformation localization. Physically, the heterogeneity of the considered microstructure induces an interaction between the formed cracks \cite{askes2000advanced}. The strains are located in a thin band, called localization band, resulting in the formation of macrocracks. Thus, the stress field at the physical point cannot be efficiently described only by the characteristics at the point but must also take into consideration its environment. Moreover, no indication about the cracking scale is included. Therefore, no information is given on the width of the localization band. The localization problem is poorly formulated mathematically as softening causes a loss of ellipticity of the differential equations describing the deformation process \cite{peerlings1996some}. The numerical solutions do not converge to physically acceptable solutions despite mesh refinements. A regularization method is, therefore, necessary to obtain a better synergy between mathematical formulation and physical phenomenon. The choice is to regularize the strain by adding the strain gradient term, and thus to use a regularized deformation tensor $\bar{\epsilon}$ which verifies the characteristic equation (see  \cite{Aifantis2020}):

\begin{equation}
\epsilon=\bar{\epsilon}-L_{c}^{2} \nabla^{2}\bar{\epsilon}
\label{eq mazars}
\end{equation}
In Marigo et al. \cite{marigo2015gradient} and Mazars and Bazant \cite{mazars1989strain} models, the failure is described by means of internal lengths $l$ (see Eq. (\ref{eq marigo})) and $L_{c}$ (Eq. (\ref{eq mazars})) which are not explicitly linked to any material parameters.

In this context, the aim of the present paper is to develop a non-local homogenized model in the elastic case by combining the second gradient theory presented in \cite{mindlin1968first} and the AEH in order to derive non-local parameters related to the microstructure.

The aim is to apply the asymptotic expansion homogenization (AEH) analysis to the probabilistic framework thanks to the introduction of the stochastic parameter $w$ in $\mathbb{R}^{3}$. For every $w=(w_{i})_{i\in\mathbb{N}}$, the convex shape $S^{\eta}(w_{i})$ is formed and, therefore, $w\mapsto S^{\eta}(w)$ is a random set in $\mathbb{R}^{3}$.  The AEH analysis is first overviewed in the context of engineering multi-scale problems, then applied in the case of a probabilistic process until revealing the emergence of a non-local term at the macroscopic scale.  Some of the main aspects of the higher order terms of the asymptotic expansion of the displacement field are also presented. A variational formulation is developed in section 3.  Theoretical developments are completed by numerical simulations for the progressive evaluation of both the local and non-local parts of the elastic energy in the stochastic case (ergodicity theory). In the three-dimensional case, the relevance of the advanced modelling is assessed for two different types of microstructures with increasing complexity in terms of morphology. Both are characterized by inclusions randomly embedded within a matrix according to a stochastic point process. The first ones are virtual generated from a fixed pattern of spherical inclusions while the second ones are real EPDM microstructures containing cavities.

\label{S:1}
 
\section{Asymptotic Expansion Homogenization}
\subsection{Generalities}
The asymptotic expansion homogenization (AEH) method was developed by Francfort \cite{francfort1983homogenization} for the case of linear thermoelasticity in periodic structures.  The AEH method has been employed to calculate the homogenized thermomechanical properties of composite materials (elastic moduli and coefficient of thermal expansion) \cite{dasgupta1996three,nasution2014thermomechanical}. The detailed numerical modeling of the mechanical behavior of composite material structures often involves high computational costs. The use of homogenization methodologies can lead to significant improvements.  For example, this technique allows the substitution of heterogeneous medium with an equivalent homogeneous medium (see Fig.  1) including second-order displacement gradients \cite{forest2000thermoelasticity, Aifantis, Aifantis2005}, thereby allowing macroscopic behavior law obtained from microstructural information.  The AEH method is both an excellent approach to solve problems involving physical phenomena in continuous media and a useful technique to study the mechanical behavior of structural components built from composite materials. The  main  advantages  of  this  methodology  lie  on  the  fact  that  (i)  it  allows a significant reduction of the problem size (number of degrees of freedom) and (ii) it has the capability to characterize stress and strain microstructural fields. In fact, unlike mean field homogenization methods, the AEH leads to specific equations that characterize these fields through the localization process.

\begin{figure}[H]
\centering
\includegraphics[scale=0.65]{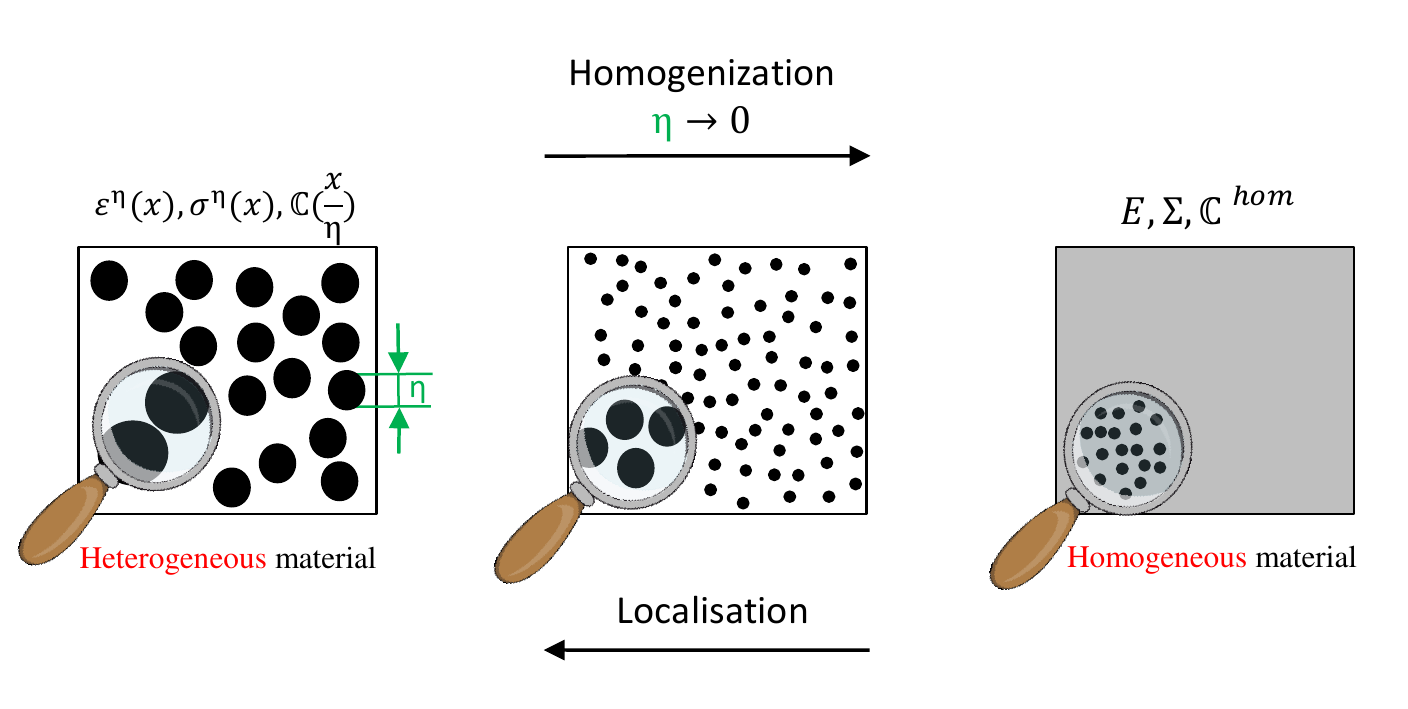}
\renewcommand{\figurename}{\bf Fig.}
\caption{Principle of asymptotic expansion homogenization method.}
\end{figure}

\subsection{Application to a probabilistic framework}
In this section, one revisits the AEH approach for random linear elastic composites defined by a  stochastic point process. One considers a heterogeneous material associated to a material body $\Omega$. Its microstructure is constituted of inclusions randomly embedded within an elastic matrix.  In classical homogenization, the original heterogeneous medium can be replaced by a homogeneous one with homogenized (so-called effective) mechanical properties provided the condition of scales separation is fulfilled. However, in real composites, the microstructural scale effects may result in specific non-local phenomena. Scale effects can be systematically analyzed by means of the higher order AEH  method. According to this approach, physical and mechanical fields in a composite are represented by multi-scale asymptotic expansions in powers of a small parameter $\eta=\frac{l}{L}$, where $l$ is the size of the representative elementary volume and $L$ is the sample/structure size. $\eta$ characterizes the heterogeneity of the composite structure. This leads to a decomposition of the final solution into macro and microcomponents. We  suppose to be in the case of periodic conditions at the scale of the REV $ Y $. Furthermore, application of the volume-integral homogenizing operator provides a link between the micro and macroscopic behaviors of the material and allows the evaluation of effective properties.

In order to separate macro and microscale components of the solution, slow $(x)$ and fast $(y)$ coordinate variables are introduced with $y=\frac{x}{\eta}$. Using both, $x$ and $y$ variables, the following chain rule of functional differentiation is used:

\begin{equation}
\frac{d (.)}{dx}=\frac{\partial (.)}{\partial x}+\frac{1}{\eta}\frac{\partial (.)}{\partial y}
\end{equation}
\subsubsection{Local problem formulation}
The local problem is described by the following equations:
\begin{equation}
\left\{ 
\begin{aligned}
&&&div(\sigma^{\eta}(x,w))\,+\,f(x,w)= 0 \\
&&&\sigma^{\eta}(x,w) =\mathbb{C}(\frac{x}{\eta},w)\,:\,\epsilon^{\eta}(x,w)\ x \in \Omega\\
&&&\epsilon^{\eta}(x,w) =sym\,(\nabla u^{\eta}(x,w))\\
&&& \mbox{\azd{Periodic boundary conditions}}
\end{aligned}
\right.
\end{equation}

where the random distribution of the inclusions is represented by the parameter $w$. This parameter is the center of inclusions randomly distributed in $\mathbb{R}^{3}$ according to a stochastic point process associated with a suitable probability space ($\Game, A, P$) (see\cite{michaille2011macroscopic}). Displacement field is denoted by $u^{\eta}(x,w)$ whereas $f(x,w)$ denotes the source terms. The latter will be neglected for all numerical simulations performed in this study. The linear elasticity tensor is noted by $\mathbb{C}(\frac{x}{\eta},w)$, which is homogeneous for each phase. The interfaces between the inclusions and the matrix are considered perfect. Thus, both the displacement field $u^{\eta}(x,w)$ and the stress vector $t^{\eta}(x,w)=\sigma^{\eta}(x,w).n$ across the interfaces with unit normal vector $n$ are continuous. Operators div and $\nabla$ denote the partial derivative ”divergence” and ”gradient”, respectively. 


The differential operations are divided into two parts: $\nabla(.)=\nabla_{x}(.)+\frac{1}{\eta}\nabla_{y}(.)$ and $div(.)=div_{x}(.)+\frac{1}{\eta}div_{y}(.)$ where the indexes $x$ and $y$ indicate that the derivatives are taken with respect to the first $(x)$ and second $(y)$ variables. The problem is therefore rewritten as:

\begin{equation}
\left\{ 
\begin{aligned}
&&&div_{x}(\sigma^{\eta}(x,w))\,+\,\frac{1}{\eta} div_{y}(\sigma^{\eta}(x,w))+ f(x,w)= 0 \\
&&&\sigma^{\eta}(x,w) =\mathbb{C}(\frac{x}{\eta},w)\,:\,\epsilon^{\eta}(x,w)\hspace{5cm} x \in \Omega\\
&&&\epsilon^{\eta}(x,w) =sym\Big((\nabla_{x}u^{\eta}(x,w))\,+\,\frac{1}{\eta}\,(\nabla_{y}u^{\eta}(x,w))\Big)\\
&&& \mbox{\azd{Periodic boundary conditions}}
\end{aligned}
\right.
\label{sys eq equ avec eta} 
\end{equation}

\subsubsection{Asymptotic expansion of the displacement field}
Following the principle of the AEH method, the displacement field can be approximated at the macroscale $\Omega$ and microscale Y with the following asymptotic expansion in $\eta$. 
\begin{equation}
\begin{split}
u^{\eta}(x,w)&= u^{0}(x,\frac{x}{\eta},w)+\eta^{1}u^{1}(x,\frac{x}{\eta},w)+\eta^{2}u^{2}(x,\frac{x}{\eta},w)+\eta^{3}u^{3}(x,\frac{x}{\eta},w)+...\\
	&= \sum_{n=0}^{n=+\infty}\eta^{n}\,u^{n}(x,\frac{x}{\eta},w)
\end{split}
\label{dep asym}
\end{equation}

where each term $u^{n}(x,\frac{x}{\eta},w)$ is a function of both variables $x$ and $y$ and depends on the stochastic point process ($w$). The first term $u^{0}$ represents the homogenized part of the solution; it changes slowly within the whole material sample. The next terms $u^{n}$, n=1,2,3,..., provide higher order corrections and describe local variations of the displacement at the scale of heterogeneities. Using the displacement asymptotic expansion in Eq. $(\ref{sys eq equ avec eta})_{3}$ leads to the strain tensor as a function of macroscopic $(x)$ and microscopic $(y)$ variables. It may be expanded in a series of powers of small (material) parameter $\eta=x/y$:

\begin{equation}
\begin{split}
\epsilon^{\eta}(x,w)&= \frac{1}{\eta}\epsilon_{y}(u^{0})+\epsilon_{x}(u^{0})+\epsilon_{y}(u^{1})+\eta[\epsilon_{x}(u^{1})+\epsilon_{y}(u^{2})]+\eta^{2}[\epsilon_{x}(u^{2})+\epsilon_{y}(u^{3})]+...\\
	&= \frac{1}{\eta}\epsilon_{y}(u^{0})+\sum_{n=0}^{n=+\infty}\eta^{n}\,\epsilon^{n}(u^{n}, u^{n+1})
\end{split}
\label{def asym}
\end{equation}

Where  $\epsilon^{n}(u^{n}, u^{n+1})=\epsilon_{x}(u^{n})+\epsilon_{y}(u^{n+1})$ with $\epsilon_{x}$ and $\epsilon_{y}$ denoting the symmetric gradients with respect to the slow and fast variables ($\epsilon_{x}=(\nabla_{x}+\nabla_{x}^{t})/2$, $\epsilon_{y}=(\nabla_{y}+\nabla_{y}^{t})/2$). In Eq. (\ref{def asym}), the strain field must be finite when $\eta\, \rightarrow\,0$. This suggests that:

\begin{equation}
\epsilon_{y}(u^{0})=0\quad and \quad {u}^{0}(x,\frac{x}{\eta},w)={U}^{0}(x)
\label{sol u 0}
\end{equation}  

We can conclude that the first term $u^{0}(x,\frac{x}{\eta},w)$ in Eq. (\ref{dep asym}) does not depend on the fast variable ($\partial u^{0}/\partial y = 0$). 

The AEH method is based on the following assumption for the stress tensor $\sigma$:

\begin{equation}
\begin{split}
\sigma^{\eta}(x,w)&= \sigma^{0}(x,\frac{x}{\eta},w)+\eta^{1}\sigma^{1}(x,\frac{x}{\eta},w)+\eta^{2}\sigma^{2}(x,\frac{x}{\eta},w)+\eta^{3}\sigma^{3}(x,\frac{x}{\eta},w)+...\\
	&= \sum_{n=0}^{n=+\infty}\eta^{n}\,\sigma^{n}(x,\frac{x}{\eta},w)
\end{split}
\label{contrainte asymp}
\end{equation}

Using Eq. (\ref{contrainte asymp}), the equilibrium equation (Eq. $(\ref{sys eq equ avec eta})_{1}$) can be written as follows:

\begin{equation}
\frac{1}{\eta}div_{y}(\sigma^{0})+\sum_{n=0}^{n=+\infty}\eta^{n}\,[div_{x}(\sigma^{n})+div_{y}(\sigma^{n+1})]+f(x,w)=0 
\end{equation}

By identification, the source term ($f(x,w)$) is associated to order 0. This leads to the following differential equations:

\begin{equation}
\left\{ 
\begin{aligned}
&&&div_{y}\,(\sigma^{0})= 0 \\
&&&div_{x}\,(\sigma^{0})+div_{y}\,(\sigma^{1})+f(x,w)=0
\end{aligned}
\right.
\end{equation}

\subsubsection{Homogenization problems and their solutions}
For successive values of $n$ (i.e for successive orders of correction), it is possible to establish hierarchical differential equations systems provided the limit of the equations exists when $\eta\, \rightarrow\,0$. In this paper, this general methodology is illustrated until the order 2 and associated solutions in terms of displacement are obtained. Then, the general solution is obtained by summation of the previous solutions.    

\textit{Problem of order 0}\\
The first hierarchical equations system (without correction) is written as follows:

\begin{equation}
\left\{ 
\begin{aligned}
&\sigma^{0}= \mathbb{C}(\frac{x}{\eta},w):(\epsilon_{x}(u^{0})+\,\epsilon_{y}(u^{1})) \\
&div_{x}(\sigma^{0})+div_{y}(\sigma^{1})\,+\,f(x,w)=\,0\\
& \mbox{\azd{Periodic boundary conditions}}
\end{aligned}
\right.
\end{equation}

where $u^{0}$ is given by Eq. (\ref{sol u 0}) and $\epsilon_{x}(u^{0})=\epsilon_{x}(U^{0})=E^{0}(x)$, $E^{0}(x)$ being the macroscopic strain in order 0.\\
   
\textit{Problem of order 1}\\
The second hierarchical equations system (first order of correction) is written as follows: 

\begin{equation}
\left\{ 
\begin{aligned}
&\sigma^{1}= \mathbb{C}(\frac{x}{\eta},w):(\epsilon_{x}(u^{1})+\,\epsilon_{y}(u^{2})) \\
&div_{x}(\sigma^{1})+div_{y}(\sigma^{2})\,=\,0\\
& \mbox{\azd{Periodic boundary conditions}}
\end{aligned}
\right.
\end{equation}

The problem associated with order 1 can be interpreted as an elasticity problem that is linear in  $E^{0}(x)$ charaterizing the loading applied. Accordingly, the solution of this problem may be written as follows:

\begin{equation}
\begin{split}
{u}^{1}(x,\frac{x}{\eta},w)&={U}^{1}(x)+{\chi}^{0}(y,w):E^{0}(x)
\end{split}
\label{sol u 1} 
\end{equation}

where $U^{1}(x)$ is a constant translation term with respect to the $x$ variable. ${\chi}^{0}(y,w)$ denotes the elastic corrector tensor or characteristic function whose volume average vanishes, $<{\chi}^{0}(y,w)>_{Y}$=0. The strain field $\epsilon^{0}$, is thus given by:

\begin{equation}
\epsilon^{0}\big(u^{0},u^{1}\big)=E^{0}(x)+\epsilon_{y}(u^{1})={\mathbb{L}}^{0}(y):E^{0}(x)
\end{equation}

where ${\mathbb{L}}^{0}(y)$, called localization tensor, is defined by: 

\begin{equation}
{\mathbb{L}}^{0}(y)={1\!\!1}+\frac{1}{2}\left\{ \frac{\partial{\chi}^{0}(y,w)}{\partial y}+\frac{\partial({\chi}^{0}(y,w))^{t}}{\partial y} \right\}
\label{L zero}
\end{equation}

where $1\!\!1$ denotes the fourth-order identity tensor.\\

\textit{Problem of order 2}\\
The third hierarchical equations system (second order of correction) is written as follows: 

\begin{equation}
\left\{ 
\begin{aligned}
&\sigma^{2}= \mathbb{C}(\frac{x}{\eta},w):(\epsilon_{x}(u^{2})+\,\epsilon_{y}(u^{3})) \\
&div_{x}(\sigma^{2})+div_{y}(\sigma^{3})\,=\,0\\
& \mbox{\azd{Periodic boundary conditions}}
\end{aligned}
\right.
\end{equation}

The displacement field in order 1 depends on macroscopic field $U^{1}(x)$ and $E^{0}(x)$ which appear in  the second order problem as loading variables. Indeed, the problem of order 2 can be interpreted as an elasticity problem that is linear in $E^{1}(x)= \epsilon_{x}(U^{1})$ and $\nabla_{x}E^{0}(x)$. The approach is thus similar to the previous one performed for the first order. The solution of the problem may be written as follows:

\begin{equation}
{u}^{2}(x,\frac{x}{\eta},w)={U}^{2}(x)+{\chi}^{0}(y,w):E^{1}(x)+{\chi}^{1}(y,w)\,\vdots\,\nabla_{x}E^{0}(x)
\label{sol u 2} 
\end{equation}

where the field $U^{2}(x)$ is a constant translation term with respect to the $x$ variable. ${\chi}^{1}(y,w)$ is a corrector tensor with a zero average value, $<{\chi}^{1}(y,w)>_{Y}$= 0. With the displacement solution, the strain field, $\epsilon^{1}\big(u^{1}, u^{2}\big)$, is given by the following expression:

\begin{equation}
\begin{split}
\epsilon^{1}(u^{1},u^{2})&=\epsilon_{x}(u^{1})+\epsilon_{y}(u^{2})={\mathbb{L}}^{0}(y):E^{1}(x)+{\mathbb{L}}^{1}(y)\vdots\nabla_{x}E^{0}(x)\\
\end{split}
\end{equation}

where ${\mathbb{L}}^{0}(y)$ is defined by Eq. (\ref{L zero}) and ${\mathbb{L}}^{1}(y)$, called localization tensor, is given by:

\begin{equation}
{\mathbb{L}}^{1}(y)=\frac{1}{2}\big({\chi}^{0}\,(y,w)\otimes{\delta}+({\chi}^{0}\,(y,w))^{t}\otimes{\delta}\big)+\frac{1}{2}\left\{\frac{\partial{\chi}^{1}(y,w)}{\partial y}+\frac{\partial({\chi}^{1}(y,w))^{t}}{\partial y} \right\}
\end{equation}

where $\otimes$ represents the tensor product operator and $\delta$ denotes the second-order identity tensor.\\

\textit{General solution}\\
The solution is obtained by summing the solution fields of problems of order 0, 1, 2. According to Eq. (\ref {sol u 0}), Eq. (\ref {sol u 1}) and Eq. (\ref {sol u 2}), the displacement fields 0, 1, and 2 have the following expressions:

\begin{equation}
\begin{aligned}
&{u}^{0}(x,\frac{x}{\eta},w)={U}^{0}(x)\\
&{u}^{1}(x,\frac{x}{\eta},w)={U}^{1}(x)+{\chi}^{0}(y,w):\nabla_{x}{U}^{0}(x) \\
&{u}^{2}(x,\frac{x}{\eta},w)={U}^{2}(x)+{\chi}^{0}(y,w):E^{1}(x)+{\chi}^{1}(y,w)\,\vdots\,\nabla_{x}E^{0}(x)
\end{aligned}
\label{solution u0u1u2 dans le sys}
\end{equation}

The whole displacement field (Eq. (\ref{dep asym})) can, therefore, be written as:

\begin{equation}
u^{\eta}(x,w)=U(x)+\eta\,{\chi}^{0}(y,w):E(x)+\eta^{2}\,{\chi}^{1}\,\vdots\,\nabla_{x}E(x)+...
\end{equation}

where the macroscopic displacement and strain fields have been introduced:

\begin{equation}
\left\{ 
\begin{aligned}
&U(x)=U^{0}(x)+\eta U^{1}(x)+\eta^{2}U^{2}(x)+...\\
&E(x)=E^{0}(x)+\eta E^{1}(x)+...
\end{aligned}
\right.
\end{equation}
  
\section{Energy method}

The behavior law at any point in a continuous medium is characterized by a strictly convex and coercive elastic potential function. Tran et al \cite{tran2012micromechanics} presented a formulation including deformation gradients. Following this idea and with moreover the introduction of the stochastic parameter $w$, the elastic energy is given by:

\begin{equation}
W^{(\eta)}(u^\eta)=W(u^0,\,u^1,\,...)=\int_{\Omega \times Y}\frac{1}{2}\left(\epsilon^{\eta}(x,w):\mathbb{C}(\frac{x}{\eta},w):\epsilon^{\eta}(x,w) \right)dx\,dy
\label{ener Tran}
\end{equation}

By replacing the expression Eq. (\ref{def asym}) of the strain tensor in the whole energy function Eq. (\ref{ener Tran}) and grouping the terms of the same power in $\eta$, the following expression is obtained:

\begin{equation}
W^{(\eta)}(u^\eta)=W(u^0,\,u^1,\,u^2\,,...)=\frac{1}{\eta^2}W^{(-2)}+\frac{1}{\eta}W^{(-1)}+W^{(0)}+\eta W^{(1)}+\eta^2 W^{(2)}+\,...
\label{enrgie asym}
\end{equation}

Where for the various power orders of $\eta$, it is possible to obtain the following set of equations:\\

\underline{Order (-2):}
\begin{equation}
\frac{1}{\eta^2}W^{(-2)}(u^0)=\frac{1}{\eta^2}\int_{\Omega \times Y} \frac{1}{2}\,\epsilon_{y}(u^0):\mathbb{C}(\frac{x}{\eta},w):\epsilon_{y}(u^0)\,dx\,dy
\label{ener ordre -2}
\end{equation}

\underline{Order (-1):}
\begin{equation}
\frac{1}{\eta}W^{(-1)}(u^0,u^1)=\frac{1}{\eta}\int_{\Omega \times Y} [\epsilon_{x}(u^0)+\epsilon_{y}(u^1)]:\mathbb{C}(\frac{x}{\eta},w):\epsilon_{y}(u^0)\,\,dx\,dy
\label{ener ordre -1}
\end{equation}

\underline{Order (0):}
\begin{equation}
\begin{split}
W^{(0)}(u^0,u^1)&=\frac{1}{2}\int_{\Omega \times Y} [\epsilon_{x}(u^0)+\epsilon_{y}(u^1)]:\mathbb{C}(\frac{x}{\eta},w):[\epsilon_{x}(u^0)+\epsilon_{y}(u^1)]\,\,dx\,dy-\int_{\Omega \times Y}f\,u^0\,\,dx\,dy\\
&with \int_{\Omega \times Y} f\,\,dx\,dy\,=\,0 
\end{split}
\label{ener ordre 0}
\end{equation}

\underline{Order (1):}
\begin{equation}
\eta W^{(1)}(u^0,u^1, u^2)=\eta \int_{\Omega \times Y} [\epsilon_{x}(u^0)+\epsilon_{y}(u^1)]:\mathbb{C}(\frac{x}{\eta},w):[\epsilon_{x}(u^1)+\epsilon_{y}(u^2)]\,dx\,dy
\label{ener ordre 1}
\end{equation}

In the present work, analytical developments have been performed until the first order of correction for the energy in order to introduce the second displacement gradient which represents the kernel of the regularized term. Now, the following subsections aim to estimate the effective properties of the heterogeneous medium. First, the common part (local part) of the energy will be detailed through theoretical developments. Then, the whole model will be  evaluated by complementary theoretical developments followed by numerical simulations for two different three-dimensional microstructures.

\subsection{Local part of the energy}
The displacement fields $u^0$ and $u^1$ are the solutions of the minimizing problem of the quadratic function $W^{(0)}$ given by Eq. (\ref{ener ordre 0}). By substituting Eq. $(\ref{solution u0u1u2 dans le sys})_{2}$ for $u^1$ in Eq. (\ref{ener ordre 0}), the energy expression $W^{(0)}$ of the order 0 for the expansion of small parameter $\eta$ is given by:

\begin{equation}
\boxed{W^{(0)}(u^0,u^1)=\frac{1}{2}\int_{\Omega \times Y} \epsilon_{x}(u^0):[1\!\!1+\nabla_{y}\chi^{0}(y,w)]:\mathbb{C}(\frac{x}{\eta},w):[1\!\!1+\nabla_{y}\chi^{0}(y,w)]:\,\epsilon_{x}(u^0)\,dx\,dy}
\end{equation}

With the macroscopic deformation defined by $\epsilon_{x}(u^{0})=\epsilon_{x}(U^{0})=E^{0}(x)$, we obtain 
\begin{equation}
W^{(0)}(u^0,u^1)=\frac{1}{2}\int_{\Omega} E^0(x):\mathbb{A}^{(0,0)}:E^0(x)\, \,dx
\label{ener ordre 0avec A00}
\end{equation}

where $\mathbb{A}^{(0,0)}$ corresponds to an elastic effective tensor, given by the following expression:
\begin{equation}
\mathbb{A}^{(0,0)}=\int_{Y} [1\!\!1+\nabla_{y}\chi^{0}(y,w)]:\mathbb{C}(\frac{x}{\eta},w):[1\!\!1+\nabla_{y}\chi^{0}(y,w)]\, dy
\end{equation}
More precisely, this tensor corresponds to a homogenized elasticity tensor. For its calculation it is only necessary to know the characteristic tensor $\chi^{0}$ which can be deduced by making several draws and then by carrying out one of a weighted average (by $\chi^{0}$) on REV $Y$. Note here that only the local part of the behavior is considered. Non-local behavior will be taken into account via the following developments.

\subsection{Non-local part of the energy}
\subsubsection{Theoretical development}
The aim of this section is to evaluate the additional constitutive properties associated with a higher order strain. The displacement fields $u^0$, $u^1$ and $u^2$ are the solutions of the minimizing problem of the quadratic function $W^{(1)}$ given by Eq. (\ref{ener ordre 1}). By substituting Eq. $(\ref{solution u0u1u2 dans le sys})_{2}$ and $(\ref{solution u0u1u2 dans le sys})_{3}$ for $u^1$ and $u^2$ in Eqs. (\ref{ener ordre 1}), the energy expression $W^{(1)}$ of the order 1 for the expansion of small parameter $\eta$ is given by:

\begin{equation}
\begin{split}
\eta W^{(1)}(u^0,u^1, u^2)&=\eta \int_{\Omega \times Y} [\epsilon_{x}(u^{0}):(1\!\!1+\nabla_{y}\chi^{0}(y,w))]:\mathbb{C}(\frac{x}{\eta},w):[(1\!\!1+\nabla_{y}\chi^{0}(y,w)):\nabla_{x}U^{1}(x)]\, dx\,dy\\
&+\eta \int_{\Omega \times Y} [\epsilon_{x}(u^{0}):(1\!\!1+\nabla_{y}\chi^{0}(y,w))]:\mathbb{C}(\frac{x}{\eta},w):\chi^{0}(y,w):\nabla_{x}\epsilon_{x}(u^{0})\, dx\,dy\\
&+\eta \int_{\Omega \times Y} [\epsilon_{x}(u^{0}):(1\!\!1+\nabla_{y}\chi^{0}(y,w))]:\mathbb{C}(\frac{x}{\eta},w):\nabla_{y}\chi^{1}(y,w)\vdots\,\nabla_{x}\epsilon_{x}(u^{0})\, dx\,dy
\end{split}
\label{ener ordre 1 forme 1}
\end{equation}

\vspace{3mm}

With E$^{0}$(x)=$\epsilon_{x}(u^{0})$ and $E^{1}(x)=\epsilon_{x}(U^{1})$ representing the macroscopic strain in order 0 and 1, respectively, Eq. (\ref{ener ordre 1 forme 1}) can be written as follows:

\begin{equation}
\begin{split}
\eta W^{(1)}(u^0,u^1, u^2)&=\eta \int_{\Omega} E^{0}(x):\mathbb{B}^{(0,1)}:E^{1}(x)\, dx\,+\eta \int_{\Omega} E^{0}(x):\mathbb{C}^{(0,0)}:\nabla_{x}E^{0}(x)\, dx\\
&+\eta \int_{\Omega} E^{0}(x):\mathbb{D}^{(0,0)}\vdots\,\nabla_{x}E^{0}(x)\, dx\,
\end{split}
\label{ener ordre 1 forme 2}
\end{equation}

where

\begin{equation}
\left\{ 
\begin{aligned}
&\mathbb{B}^{(0,1)}=\int_{Y}[(1\!\!1+\nabla_{y}\chi^{0}(y,w))]:\mathbb{C}(\frac{x}{\eta},w):[(1\!\!1+\nabla_{y}\chi^{0}(y,w))]\,dy\\
&\mathbb{C}^{(0,0)}=\int_{Y}[1\!\!1+\nabla_{y}\chi^{0}(y,w))]:\mathbb{C}(\frac{x}{\eta},w):\chi^{0}(y,w)\,dy\\
&\mathbb{D}^{(0,0)}=\int_{Y} [1\!\!1+\nabla_{y}\chi^{0}(y,w))]:\mathbb{C}(\frac{x}{\eta},w):\nabla_{y}\chi^{1}(y,w)\,dy
\end{aligned}
\right.
\label{B01 C00 D00}
\end{equation}

Eq. (\ref{ener ordre 1 forme 2}) involves the first strain gradient $\nabla_{x}E^{0}$ which is equivalent to the second-order displacement gradient $\nabla\nabla_{x}U^{0}$ we wanted to make appear. In other words, it denotes the main/key of the regularized term. Also, three homogenized elasticity tensors $(\mathbb{B}^{(0,1)},\,\mathbb{C}^{(0,0)}$ and $\mathbb{D}^{(0,0)})$ naturally emerge in Eq. (\ref{ener ordre 1 forme 2}). They are given by Eq. (\ref{B01 C00 D00}) and they have the order 4, 3 and 5, respectively. It is interesting to note that through the characteristic tensors ($\chi^{0}(y,w)$ and $\chi^{1}(y,w)$), the random distribution is taken into account during the scale transition so as to preserve statistical information. In addition, a non-local effect after homogenization is evidenced through the presence of the gradients of the characteristic tensors in the expressions of the homogenized tensors.

The accuracy of the proposed model was assessed by computing the whole energy and comparing its predictions with the classical bounds. According to Eq. (\ref{enrgie asym}) until order 1 and with $\frac{1}{\eta^{2}}W^{(-2)}=\frac{1}{\eta}W^{(-1)}=0$, the whole energy is given by:

\begin{equation}
W^{(\eta)}(u^{\eta})= W^{(0)}+\eta W^{(1)} 
\label{ener forme finale}
\end{equation} 

In order to compute $E^{0}(x)$ and $E^{1}(x)$, full-field simulations over too different morphological representative elementary volumes (MREV) were performed. They are noted MREV0 and MREV1, respectively. This required to determine two characteristic lengths $l_{0}$ and $l_{1}$ defining the size of both these volumes. An accurate way of doing this is to do a statistical analysis through the covariogram method, the latter is brifely recalled below.\\

The covariance is a very useful characteristics for the description of the size, shape and spatial distribution of a given particle. Covariance $C(x, x+h)$ is defined by the probability  $P$ for two points, separated by the vector $h,$ to belong to the same stationary random set $B$ (see\cite{azdine2015catching, jeulin2000random, lantuejoul2013geostatistical, torquato2013random} and \cite{serra1982image}):

\begin{equation}
C(x,x+h)=P\lbrace x\in B,x+h\in B\rbrace
\end{equation}
Many different data can be estimated from this statistical tool. Although a covariogram has many properties, only the correlation distance $Dc(.)$ and the outdistance repulsion $Dr(.)$ were used in this study. The first intersection between $C(h)$ and the asymptote corresponds to the correlation length. It is defined by:
\begin{figure}[H]
\centering
\includegraphics[scale=0.7]{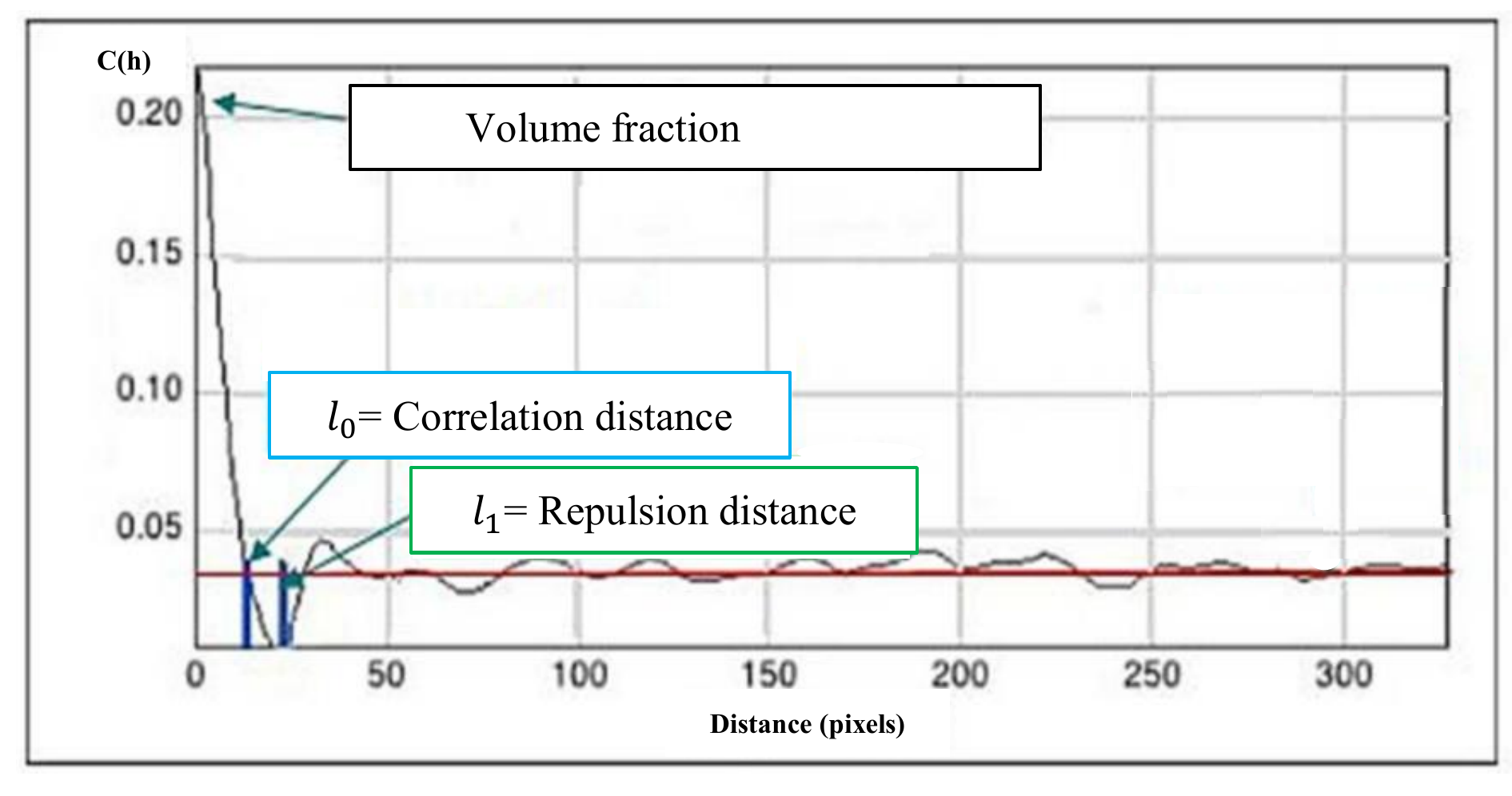}
\renewcommand{\figurename}{\bf Fig.}
\caption{Example of covariogram for a 3D heterogeneous medium and identification of characteristic lenghts $"l_0"$ and $"l_1"$.}
\end{figure}

\begin{equation}
Dc(z)=\min_{h\in \mathbb{R}^{*} }\lbrace Cz(h)- p^{2} =0 \rbrace
\end{equation}

for a given direction $z$. The quantity $p$ denotes the volume fraction of the phase B.This distance represents the maximal distance of statistical influence of the inclusion phase. It provides information about the minimal size of the domain over which a volume is statistically representative. Beyond this distance, additional statistical information is negligible (see \cite{azdine2015catching}). Thus, this size is the material first characteristic length (see Fig. 2) used to define the size of MREV0. It is noted $"l_0"$.

The second intersection between $C(h)$ and the asymptote corresponds to an outdistance of repulsion. It corresponds to the statistical average distance between two inclusions. In clustering situations, $Dr(.)$ gives an estimate of the statistical average distance between two clusters in each direction. Thus, similar to the correlation length, this distance allowed us to estimate the second characteristic length named $"l_1"$ (see Fig. 2), used to generate the second MREV1. More precisely, this distance presents the range of the first volume (MREV0).

To summarize, the main steps used to compute the whole energy (Eq. (\ref{ener forme finale})) were the following:

%
%
%
%
%
%
%

\begin{figure}[H]
\centering
\includegraphics[scale=0.7]{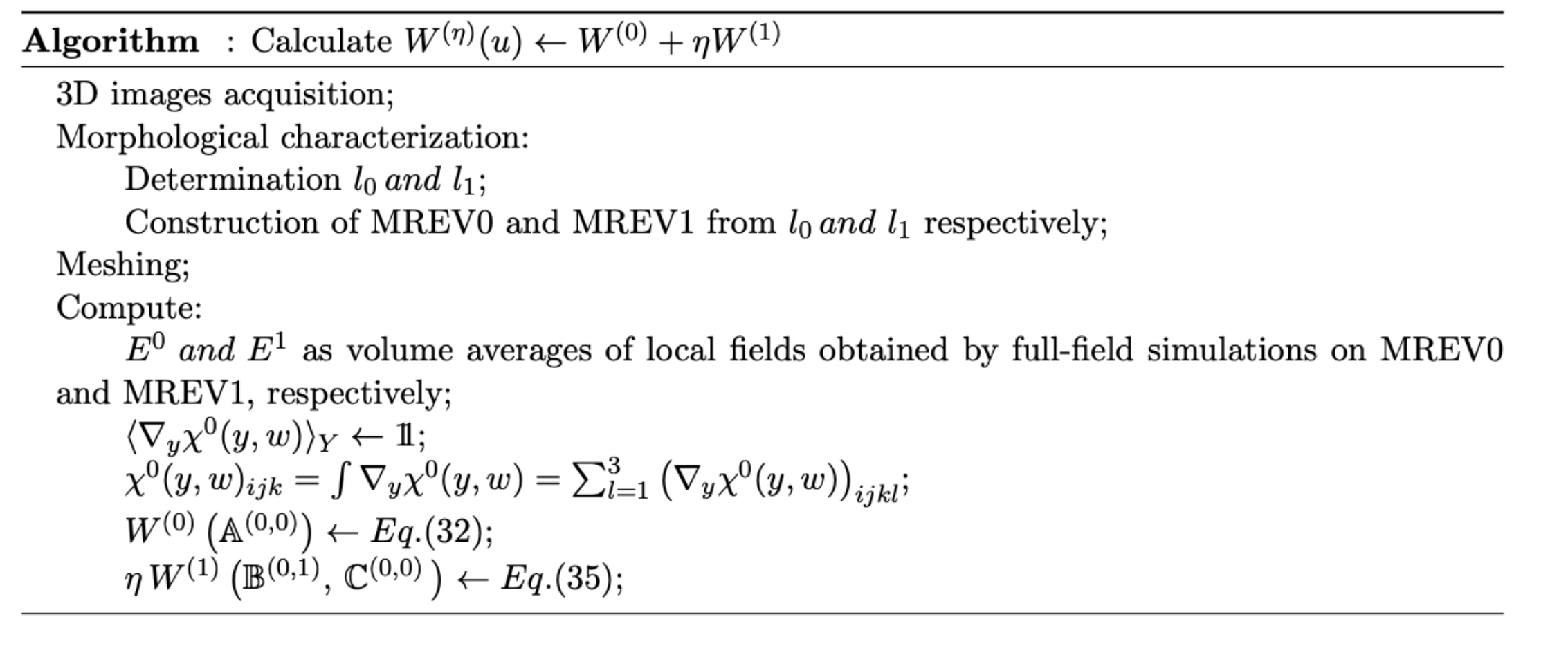}
\renewcommand{\figurename}{\bf Fig.}
\end{figure}

For a first approach, $\mathbb{D}^{(0,0)}$ is not yet taken into account in the computation. 

\subsubsection{3D Numerical simulations}

Two different types of 3D microstructures were used as supports to test the proposed model. They were chosen as examples of two-phase heterogeneous materials with an elastic matrix containing a random distribution of inclusions. The loading was a unit unit uniaxial tension imposed by periodic boundary conditions. The first microstructures are virtual while the second ones correspond to unfilled EPDM microstructures obtained by High-Resolution X-ray Computed Tomography (HRXCT) \citep{orlov2009simulation} \cite{buffiere2010situ} at different times of the decompression stage after hydrogen exposure of the material \citep{castagnet2018situ}. For both microstructure types and every inclusion volume fraction investigated, the Young$'$s modulus of the inclusions was equal to 100 GPa while it was equal to 1 GPa for the matrix (contrast 100). The Poisson ratio for both constituents was 0.3. 

The virtual microstructures were generated from a simple pattern composed of a big inclusion circled by six identical small inclusions. This pattern was the same for a fixed inclusions volume fraction and only the size of both types of particles was homothetically enlarged to adjust to the desired volume fraction. For each volume fraction, $10$ realizations were generated with this pattern by a stochastic process assumed to be stationary and ergodic.  For each realization, the characteristic lengths $l_0$ and $l_1$ were determined thanks to the covariogram analysis. The arithmetic average over the $10$ resulting values of $l_0$, respectively $l_1$, was used to define the size of MREV0, respectively MREV1. Both volumes were then numerically generated and meshed with 4-node linear tetrahedral elements. Finally, the tensors $E^0$ and $E^1$ were computed from full-field simulations on MREV0 and MREV1 and the energy was derived according to the methodology exposed in section 3.2.1. Fig. 3 provides an example of realization for an inclusion volume fraction of 0.01     as well as images of MREV0 and MREV1.

At each time of the decompression, the volume fraction but also the morphology of the inclusions in the real EPDM microstructures are different contrarily to the virtual microstructures for which it is identical for every volume fraction. This introduces an additional complexity. For a given volume fraction, i.e. for a given image acquired at a specific time, the characteristics lengths $l_0$ and $l_1$ were deduced from the covariogram analysis. Then, 15 realizations of MREV0 and 15 realizations of MREV1 were randomly extracted from this HRXCT image. They were meshed by converting voxels into hexahedral elements. The tensor $E^0$, respectively $E^1$, was calculated as the ensemble average of $E^0$, respectively $E^1$, on the number of realizations of MREV0, respectively MREV1. Fig. 4 presents a 3D image of EPDM for a cavity volume fraction of 0.045 as well as an example of realization of each volume MREV0 and MREV1. 

All the FE full-field simulations were performed with an in-house finite element solver FoXtroT \citep{FoxFE}. Microstructures in Fig. 3.a and Fig. 4.a were meshed with 1.291.025 and 158.340.421 elements, respectively.

\begin{figure}[H]
\centering
\includegraphics[scale=0.45]{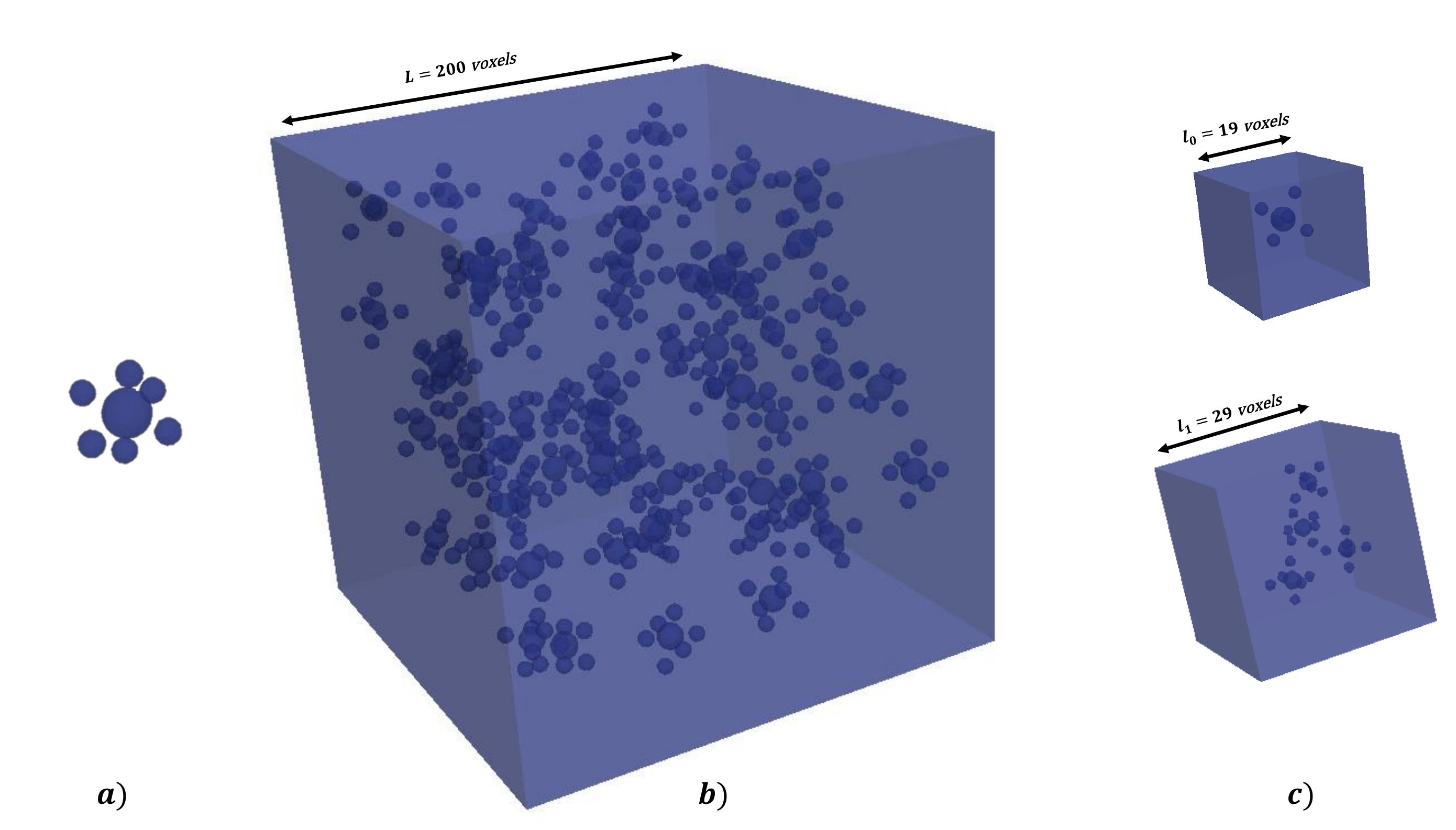}
\renewcommand{\figurename}{\bf Fig.}
\caption{Virtual microstructure a): Morphological pattern, the radius of the big, respectively small, inclusion is 1, respectively 0.5, voxel, b): Example of realization for an inclusion volume fraction of 0.01, c): Corresponding volumes MREV0 and MREV1.}
\end{figure}

\begin{figure}[H]
\centering
\includegraphics[scale=0.50]{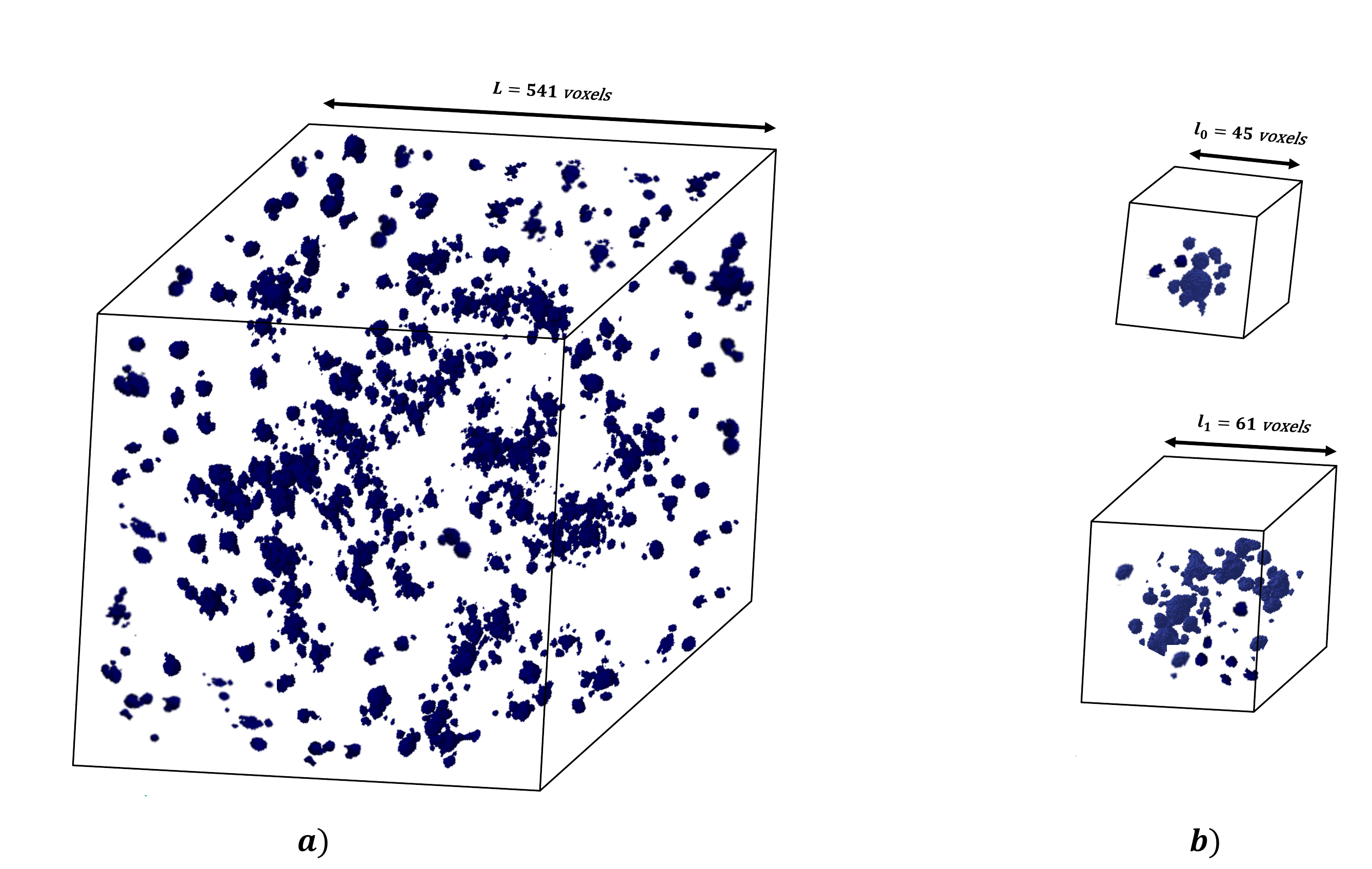}
\renewcommand{\figurename}{\bf Fig.}
\caption{Real EPDM microstruture containing cavities a): 3D image for a cavity volume fraction of 0.045 b): Examples of realization of volumes  MREV0 and MREV1 extracted from the previous 3D image.}
\end{figure}

\vspace{1.5cm}

Fig.5 represents the elastic energy given by Eq. (\ref{ener forme finale}) as a function of the inclusion volume fraction for the virtual microstructure (Fig. 3.a) and the EPDM one (Fig 4.a). The results are compared to Voigt and Reuss bounds. The values obtained by full-field simulations on the whole microstructure are also reported in Fig. 5.a for the virtual material. It was not possible to do the same for the EPDM microstructures due the size of the HRXCT images and consecutive high number of nodes to consider. For both microstructures the simulated energy values are between the bounds. They fall very close to the Reuss bound when decreasing the inclusion volume fraction. The results can be explained by the fact that these computational results were obtained for the first higher-order displacement corrector ($\eta^{1}$) only leading to the so-called underestimation.  \azd{They constitute a first validation of the model showing that the scale transition approach respects the classical limits used in homogenization.}

\begin{figure}[H]
\centering
\includegraphics[scale=0.6]{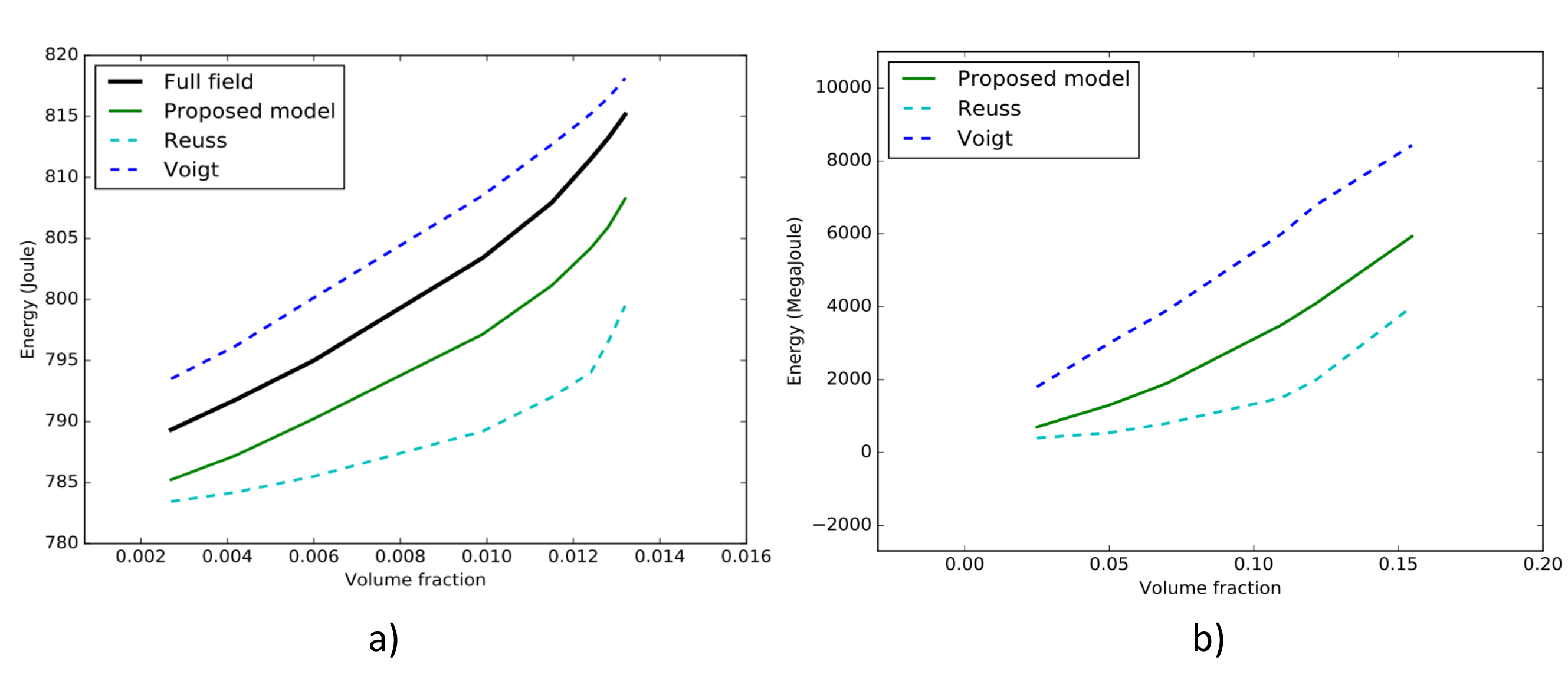}
\renewcommand{\figurename}{\bf Fig.}
\caption{Voigt-Reuss-Hill spindle. a): Virtual microstructure, b): Real EPDM Material}
\end{figure}

\section{Conclusions} \label{Conclusions}

In this paper, we developed a non-local (second-gradient) homogenenized model for three-dimensional composites in the framework of ergodic linear elastic two-phases (matrix-inclusions) random microstructures. This was done by using the asymptotic expansion homogenization (AEH) in order to derive non-local parameters related to the microstructure. The formal mathematical formulation of the AEH was detailed. The resulting sets of homogenization problems and their solutions were established. To the best of our knowledge, this was the first application of the AEH for random heterogeneous media defined by a stochastic point process. Analytical development of the homogenized elastic energy until the first order of correction makes appear the second gradient of the displacement field which represents the kernel of the regularized term. In addition, the close-form expression of the non-local part of the energy involves homogenized elasticity tensors explicitly dependent on the characteristic tensors. Thus, the advanced modeling allows to obtain a non-local macroscopic model in which statistical data are preserved. The model was tested through an extensive numerical study performed in three-dimensional cases. 

Two full-field calculations in order to compute the two strain tensors $E^{0}(x)$ and $E{^1}(x)$, which were identified respectively in order 0 and 1 of $\eta$, were required. So, in the future, we would like to obtain an analytical formulation of the model by using $\Gamma$-Convergence \citep{dal1993introduction, nait2014volumic} to avoid any full-field computation. This would allow using the model for structure calculations.

Although developed for linear elasticity, the present approach could be also extended to inelastic and/or non-linear problems in general, more particurlarly for the damage modelling \citep{bourdin2000numerical}, \citep{pham2011gradient} and \citep{xia2017phase}.

\azd{To conclude, this model is built in order to respond to the problem of materials with high property gradients. For these materials the modeling approaches modeling's are limited. And this is even more true for scale transition approaches based on the scale separation hypothesis that  makes difficult the modelling of the non-locality. The model will therefore be usable for porous materials \citep{Milhet2018} with a two scales cavity distribution \citep{castagnet2018situ, Castagnet2019}, depending on the time  \citep{KaneDiallo2016}  , or even foundry materials and additive manufactured materials. The comparison of the model results for these different classes of materials will be made in a future study.}

\section*{Acknowledgments} \label{Acknowledgments}

This work was partially funded by the French Government programs "Investissements d'Avenir" LABEX INTERACTIFS (reference ANR-11-LABX-0017-01) and EQUIPEX GAP (reference ANR-11-EQPX-0018).

Computations were performed on the supercomputer facilities
of the Mesocentre de calcul SPIN Poitou-Charentes.



\bibliographystyle{model1-num-names}
\bibliography{sampleAz.bib}

\end{document}